# CQED quantum tomography of a microwave range

## G.P. Miroshnichenko
gpmirosh@gmail.com

ITMO University, Saint Petersburg, 197101, Russian Federation

The protocol of measurement the probability density $P_\rho(\chi)$ of a rotated quadrature $\chi$ of microwave mode in an arbitrary quantum state is presented and numerically investigated. The protocol is adapted to the measuring procedure used in the CQED experiments. The method is based on measurement of integrated probability $P(m,n)$ of detection of $m$ atoms - probes in the top working state from $n$ atoms flying via the cavity. Connection of random variables $m = m(\varphi)$ and $\chi = \chi(\varphi)$ is found, when $\varphi$ is a qudrature phase. Connection of the tomogram $P(m,n)$ and the tomogram $P_\rho(\chi)$ is found. The protocol prescribes the rule of choice of problem parameters, such as interaction time, and the number of atoms in series $n$. The method contains a single fitting parameter that is selected in the process of numerical simulation of the quadrature distribution in a coherent state. Quality selection is determined by the Kolmogorov test. The result of the calculation protocol is a convolution $P_\rho(\chi)$ with the instrumental function of the method. The parameters of the instrumental function are shown. Numerical simulations performed for typical numerical values of parameters of the experimental setup, shows the effectiveness of the proposed method.

## 1. Introduction

The foundation of a quantum tomography of optical range is laid in works [1 - 4]. The review of the first works can be found in [5]. In these works the method of a tomography is used for reconstruction of quasidistributions of optical mode on the basis of measurements of the tomogram – distribution of the rotated quadrature components $\chi(\varphi)$ at different angles of rotation

$$\chi(\varphi) = \frac{1}{\sqrt{2}}\left(a^\dagger \exp(i\varphi) + a \exp(-i\varphi)\right). \qquad (1)$$

Here $a^\dagger$, $a$ - creation and annihilation operators of the photon of the studied mode. Homodyne detection method is the basis of optical tomography protocol. This method is effectively applied to Gaussian states of the optical mode. Quasidistributions $W(\alpha,s)$ of the studied mode are connected with density matrix the following formulas

$$W(\alpha,s) = Sp(\rho T(\alpha,s)),$$

$$T(\alpha,s) = \frac{1}{\pi}\int \exp(\alpha\xi^* - \alpha^*\xi)\exp\left(s\frac{|\xi|^2}{2}\right)D(\xi)d^2\xi,$$

$$D(\xi) = \exp(a^\dagger\xi - a\xi^*).$$

Here $T(\alpha,s)$ - a characteristic function, $D(\xi)$ - the shift operator in the complex plane, $s$ - the order parameter. At $s = -1, 0, 1$ quasidistribution correspond to known functions $Q(\alpha), W(\alpha), P(\alpha)$ on the complex plane $\alpha$. Connection of the tomogram $P_{\rho_0}(\chi(\varphi))$ (distribution of the quadrature in the state $\rho_0$) with the Wigner function $W(\alpha)$ is determined by the formula

$$P_{\rho_0}(\chi(\varphi)) = \langle \chi(\varphi)|\rho_0|\chi(\varphi)\rangle =$$

$$= \int_{-\infty}^{\infty} W(\chi(\varphi)\cdot\cos(\varphi) + p\cdot\sin(\varphi), -\chi(\varphi)\cdot\sin(\varphi) + p\cdot\cos(\varphi))dp,$$

$$W(x,y) =$$

$$= \int_0^\pi \frac{d\varphi}{\pi}\int_{-\infty}^\infty d\chi'(\varphi)P_{\rho_0}(\chi'(\varphi))\int_{-\infty}^\infty \frac{dk}{4}|k|\exp(ik(\chi'(\varphi) - x\cos(\varphi) + y\sin(\varphi))).$$

Here the inverse Radon transformation is executed. This transformation is singular, and the difficulty of its application is discussed in [2 - 4]. In [6 – 8] noted that the tomogram can be primary information, which can be used to calculate various physical characteristics, to verify the accuracy of uncertainty relations [9], to verify the degree of purity of quantum states [10].

Microwave range represents essential interest for quantum informatics [11, 12]. Quantum properties of microsystems can be observed if decoherence, destroying quantum superposition of states is weakened [13]. Currently, experiments with microwave photons and qubits are carried in two directions, called circuit quantum electrodynamics (cQED) and cavity quantum electrodynamics (CQED). Experimenters working in the field cEED, use special methods of measurement [14 - 22] and the protocols of quantum tomography [23 – 25]. CQED studies the properties of atoms coupled to discrete photon modes in high- Q cavities. Methods of preparation of various quantum states of photons, such as Fock, squeezed, Schrödinger's cat states [11] proposed and realized in CQED experiments. Quantum states of the mode described in terms of quasiprobabilities in phase space. In the microwave range, in CQED experiments the optical homodyne method is impossible due to the lack of the corresponding quantum detectors of microwave range. In CQED experiments, information about the state of mode is received by the atoms-probes, passed through the cavity. The states of probes emerging from the cavity are measured with help of ionization chambers. There are several methods of extracting quasiprobabilities, in particular of the Wigner function of mode, from the statistics of flying probes [26 – 29]. Of particular interest is a method of quantum tomography of the Wigner function, proposed in [30]. On this method experiments on reconstruction of the Wigner function are made [31 – 33]. In the method [30] it is shown that the Wigner function $W(\alpha)$ is extracted from the experiment at the point $\alpha$ on the complex plane. Wigner function associated with a difference of probabilities to detect the emitted probe at the top $P_e$ and bottom $P_g$ working state according to the formula

$$P_e - P_g = Sp\left(D(\alpha)\rho D^{-1}(\alpha)\exp(i\pi a^\dagger a)\right) = \frac{1}{2}W(-\alpha). \qquad (2)$$

Here $\rho$ is unknown density matrix of mode, $D(\alpha)$ is a shift operator shifting operators $a$, $a^\dagger$ on complex amplitude $\alpha$. Such shift is provided with the additional coherent classical microwave field acting on the cavity with the studied mode. Atomic Ramsey interferometer is used to obtain the operator of parity $\exp(i\pi a^\dagger a)$ under the sign of averaging. The interferometer operates at microwave range and measures the phase difference of atom-probe states superposition. Atom-probe has a dispersive interaction with the mode, due to the large detuning from resonance. The phase shift at a single photon in the interferometer must be equal $\pi$. As noted in [31], the phase shift dependent on operator $a^\dagger a$ nonlinearly, so the direct use of formulas (2) to measure the Wigner function is difficult. As an alternative in CQED researchers apply the protocol of a photon number tomography [31, 34 – 36].

In the proposed work the protocol of quadrature distribution measurement of microwave mode in CQED experiment is considered. The proposed method is based on a simpler experimental setup compared to [31]. It is shown that the measurement of the tomogram is possible on the experimental setup, does not requiring the fine tuned atomic interferometer and additional classical source of coherent radiation. Installation of a quadrature phase $\varphi$ is carried out by means of transformation of an atomic basis after a departure of atoms - probes from the cavity. The primary measured random variable is the number $m$ of atoms - probes detected in top state in a sample of $n$ atoms-probes passed through the cavity for a given phase difference $\varphi$. The protocol is based on the relation of the measured random variable $m(\varphi)$, with the calculated random value $x(\varphi)$. This relation is found in the work. The protocol of measurement gives not the tomogram $P_{\rho_0}(\chi(\varphi))$, but its convolution with the instrumental function of the method. This instrumental function is Gaussian with the dispersion found in the work and a zero average. For this reason the last action of the protocol is procedure of deconvolution. Numerical simulations performed for the numerical values of the parameters of the experimental setup [31], shows the effectiveness of the proposed method.

## 2. Conditional reduced density matrix in the CQED experiment

In the current work, we proposed a protocol of measuring the probability density of the quadrature operator (1). Quadrature distribution $P_{\rho_0}(\chi(\varphi))$ in $\rho_0$ state is connected with Glauber's function $P_0(\beta)$ [37, 38] according to a ratio

$$P_\beta(\chi(\varphi)) = |\langle \chi(\varphi)|\beta\rangle|^2 = \frac{1}{\sqrt{2\sigma_\beta \pi}} \exp\left(-\frac{\left(\chi - \sqrt{2}|\beta|\cos(\Phi - \varphi)\right)^2}{2\sigma_\beta}\right), \quad (3)$$

$$P_{\rho_0}(\chi(\varphi)) = \langle \chi(\varphi)|\rho_0|\chi(\varphi)\rangle = \iint P_0(\beta) P_\beta(\chi(\varphi)) d^2\beta. \quad (4)$$

Here $|\chi(\varphi)\rangle$ - the eigenvector belonging to the continuous spectrum of the operator $\chi(\varphi)$, $|\beta\rangle$ is a coherent state of mode

$$|\beta\rangle = \exp\left(-\frac{1}{2}|\beta|^2\right) \sum_{k=0} \frac{\beta^k}{\sqrt{k!}} |k\rangle,$$
$$\beta = |\beta|\exp(i\Phi), \quad (5)$$

$P_\beta(\chi(\varphi))$ - quadrature distribution in coherent state, quadrature dispersion is

$$\sigma_\beta = \frac{1}{2}.$$

Our purpose consists in search of the protocol of measurements of probability density $P_{\rho_0}(\chi(\varphi))$ in any mixed state $\rho_0$ of microwave cavity mode. We will focus on the parameters of the experimental setup used in [11, 27, 31, 32]. Due to the high quality factor ($Q \approx 10^{10}$) of the cavity at times of several hundred milliseconds is possible to neglect relaxation processes inside the cavity. We believe that the atom-probe, flying through the cavity, interacts only with the studied microwave mode. The scheme of experiment in which the proposed protocol is realized, is shown in Fig. 1. The experiment consists of sending through the cavity with microwave studied mode a sequence of $n$ the probe atoms (b) in a lower working state. The atoms fly into the cavity (c) alone. The pulse of classical microwave field (e) acts on the atoms - probes who flying from the cavity. As a result of this action the atomic basis will be transformed and the phase $\varphi$, necessary for measurement of a quadrature, is entered. Quantum measurement of the energy state (f) is performed on each atom at the exit of the cavity. Fig. 2 is the timing diagram of the protocol events. We define the interaction Hamiltonian of the microwave mode with the atom-probe in a resonant representation

$$V_{af} = \lambda\left(|0\rangle\langle 1|a^\dagger + |1\rangle\langle 0|a\right).$$

Here $|0\rangle$ is a vector of the lower working level of the atom-probe, $|1\rangle$ is the vector of the upper working level. The cavity mode is tuned in exact resonance with the working atomic transition. The interaction Hamiltonian of the atom-probe with a classical field pulse converts the atomic basis before measurement and is given in the resonant representation

$$V_a = f\left(\exp(-i\varphi)|0\rangle\langle 1| + \exp(i\varphi)|1\rangle\langle 0|\right).$$

The evolution operator on an interval $t_{k+1} - \Delta t - \tau \leq t < t_{k+1} - \Delta t$ in the resonance representation has the form

$$U_{af} = \exp(-iV_{af}\tau).$$

The evolution operator on an interval $t_{k+1} - \Delta t \leq t < t_{k+1}$ has the form

$$U_a = \exp(-iV_a \Delta t).$$

Here parameters $\Delta t$ and $f$ are connected by the condition

$$f \cdot \Delta t = \frac{\pi}{4}.$$

Density matrix of the field mode and the atom-probe at the time of measurement $t_{k+1}$ has the form

$$\rho_{af}(t_{k+1}) = U_a U_{af} \rho_{af}(t_k) U_{af}^\dagger U_a^\dagger.$$

Here atom – field density matrix after a measurement at time $t_k$ has the form

$$\rho_{af}(t_k) = |0\rangle\langle 0|\rho_f(t_k).$$

Here $|0\rangle\langle 0|$ is the atomic density matrix of the atom entering the cavity in the lower state, $\rho_f(t_k)$ is the reduced field density matrix after the measurement at the time $t_k$. The recurrence relation between the reduced density matrix $\rho_f(t_k)$ of the mode at the time $t_k$ and the density matrix at the time $t_{k+1}$, $k = 0,1,....n$ has the form in a resonant representation

$$\rho_f(t_{k+1}) = \frac{K_{0,0}\rho_f(t_k)(K^\dagger)_{0,0}}{P_{0,0}(t_{k+1})},$$

$$\rho_f(t_{k+1}) = \frac{K_{1,0}\rho_f(t_k)(K^\dagger)_{0,1}}{P_{1,0}(t_{k+1})}. \tag{6}$$

Here $\rho_f(t_0) = \rho_0$ is the initial density matrix. The a priori probability of finding an atom probe in the upper $P_{1,0}(t_{k+1})$ and lower $P_{0,0}(t_{k+1})$ energy state at the time of measurement $t_{k+1}$ is defined by the formula

$$P_{0,0}(t_{k+1}) = Sp_f\left(K_{0,0}\rho_f(t_k)(K^\dagger)_{0,0}\right),$$

$$P_{1,0}(t_{k+1}) = Sp_f\left(K_{1,0}\rho_f(t_k)(K^\dagger)_{0,1}\right). \tag{7}$$

Here, it is assumed that the probe atom fly into the cavity in the lower working state. Kraus's operators for the circuit shown in Fig. 1, can be written as

$$K_{00} = \frac{1}{\sqrt{2}}\left(\cos(\lambda\tau\sqrt{a^\dagger a}) - \exp(-i\varphi)a\frac{\sin(\lambda\tau\sqrt{a^\dagger a})}{\sqrt{a^\dagger a}}\right),$$

$$(K^\dagger)_{00} = \frac{1}{\sqrt{2}}\left(\cos(\lambda\tau\sqrt{a^\dagger a}) - \exp(i\varphi)\frac{\sin(\lambda\tau\sqrt{a^\dagger a})}{\sqrt{a^\dagger a}}a^\dagger\right),$$

$$K_{10} = \frac{1}{\sqrt{2}}\left(\cos(\lambda\tau\sqrt{a^\dagger a}) + \exp(-i\varphi)a\frac{\sin(\lambda\tau\sqrt{a^\dagger a})}{\sqrt{a^\dagger a}}\right),$$

$$(K^\dagger)_{01} = \frac{1}{\sqrt{2}}\left(\cos(\lambda\tau\sqrt{a^\dagger a}) + \exp(i\varphi)\frac{\sin(\lambda\tau\sqrt{a^\dagger a})}{\sqrt{a^\dagger a}}a^\dagger\right).$$

Kraus's operators reduce density matrix at the time of quantum measurement [39 – 41].

## 3. Integrated probability of detection

Directly measured random variable for each series from $n$ atoms is the number $m$ - amount of the atoms - probes found in the top working state. Experiment is repeated $N$ time for receiving integrated distribution $P_{\rho_0}(n;m)$ of the random variable $m$ changing in an interval $0 \leq m \leq n$. Our goal is to find the relationship between the measured random variable $m$ and the required random variable $\chi(\varphi)$, and as a result, to find the relation between the measured distribution $P_{\rho_0}(n;m)$ and the required distribution $P_{\rho_0}(\chi(\varphi))$. We will receive a theoretical formula for integrated probability $P_{\rho_0}(n;m)$

$$P_{\rho_0}(n;m) = \sum_{j=1} P_{\rho_0}^j(n;m),$$

$$P_{\rho_0}^{j}(n;m) = Sp_f\left(\{...K_{10}...K_{00}...\}_j \rho_0 \{...K^{\dagger}_{00}...K^{\dagger}_{01}...\}_j\right). \tag{8}$$

Here the expression $\{...K_{10}...K_{00}...\}_j$ means the $j$-th $\left(1 \leq j \leq \dfrac{n!}{(n-m)!m!}\right)$ arrangement of $n$ Kraus operators, among which there are $m$ operators $K_{10}$. It is easy to show that $P_{\rho_0}^{j}(n;m)$ is calculated as the product of conditional a priori probabilities of detection of the next atom in the upper state at the time $t_{k+1}$, $k=0,1,....n$. Conditionality means that for the previous measurement at time $t_k$ the normalized atomic - field density matrix was reduced and becomes equal to $\rho_f(t_k)$

$$P_{\rho_0}^{j}(n;m) = ... \cdot P_{1,0}(t_{k+1})\cdot ...P_{0,0}(t_{p+1})\cdot .... \tag{9}$$

The probabilities $P_{1,0}(t_{k+1})$, $P_{0,0}(t_{k+1})$ are calculated by the formulas (6), (7). Generally probabilities $P_{\rho_0}^{j}(n;m)$ differ from each other due to the effect of a reduction, and depend on a way of arrangement $j$. Let us rewrite formula (8), through Glauber's function $P_0(\beta)$ (4) for the studied matrix $\rho_0$ of a microwave mode

$$P_{\rho_0}^{j}(n;m) = \iint d^2\beta\, P_0(\beta) Sp_f\left(\{...K_{10}...K_{00}...\}_j |\beta\rangle\langle\beta| \{...K^{\dagger}_{00}...K^{\dagger}_{01}...\}_j\right) =$$
$$= \iint P_0(\beta) P_{\beta}^{j}(n;m) d^2\beta.$$

The probability $P_{\beta}^{j}(n;m)$, according to (9), can be written as the product of

$$P_{\beta}^{j}(n;m) = ... \cdot P_{1,0}^{\beta}(t_{k+1})\cdot ...P_{0,0}^{\beta}(t_{p+1})\cdot ... .$$

Here the index $\beta$ indicates that the initial condition for the calculation $P_{\beta}^{j}(n;m)$ was the coherent state $|\beta\rangle$ (5). We obtain a formula for the a priori probability $P_{1,0}^{\beta}(t_1)$ on the first measurement. In this case the mode density matrix at the time $t_0$ is equal to an initial matrix

$$\rho_0 = |\beta\rangle\langle\beta|.$$

The required formula has the form

$$P_{1,0}^{\beta}(t_1) = Sp_f\left(K_{1,0}|\beta\rangle\langle\beta|(K^{\dagger})_{0,1}\right) = \frac{1}{2}\left(1 + 2\lambda\tau|\beta|\Gamma\cos(\Phi-\varphi)\right). \tag{10}$$

Here the parameter $\Gamma$ is

$$\Gamma = \sum_{m=0}^{\infty}\cos(\lambda\tau\sqrt{n})\frac{\sin(\lambda\tau\sqrt{n+1})}{\lambda\tau\sqrt{n+1}}\exp(-|\beta|^2)\frac{|\beta|^{2n}}{n!}.$$

Given that

$$|\beta|\lambda\tau \ll 1, \tag{11}$$

$\Gamma$ has the form

$$\Gamma \approx 1 - (\lambda\tau)^2\left(\frac{2}{3}|\beta|^2 + \frac{1}{6}\right).$$

In these formulas, the parameter $\lambda = 2\pi\Omega$ is the interaction constant, where $\Omega$ is the Rabi frequency for a single photon. The graphs of dependence the probability $P_{1,0}^{\beta}(t_1)$ (10) on $|\beta|$ for different values of the phase $\varphi$ and $\Phi$ are presented in Fig. 3. As can be seen from Fig.3, under the condition (11) plots close to linear on the variable $|\beta|$. The presence of the linear part of graphs on the variable $|\beta|$ is essential for the implementation of the proposed protocol. We will study dependence of conditional a priori probability of detecting the atom in the upper state on the number $1 \leq k \leq n$ of atom, passing through the cavity. Consider the sequence of $n$ passing through the cavity the atoms-probes. Quantum measurement of an energy state is carried out over each atom emitted. We assume an ideal detection process. After each measurement density matrix is reduced according to random results of a measurement. Graphs of dependence of a priori probability $P_{1,0}^{\beta}(t_k)$ on number $k$ of atom - probe are shown in Fig. 4. Graphs of Fig. 4 are simulated by means of a stochastic recurrence relation (7). The stochastic Monte-Carlo method is used in calculation, the density matrix reduction on each measurement is considered. Condition (11) holds for all graphs of Fig. 4. As follows from the formula (10), $P_{1,0}^{\beta}(t_1)$ is different from $\frac{1}{2}$ to $|\beta|\lambda\tau$ value. Under the condition (11) graphs in Fig. 4 can be approximated by straight lines. These straight lines are parallel to axis $k$, and are defined by average value of probability $\overline{P}_{10}$

$$\overline{P}_{10} = \frac{1}{2}\left(1 + \nu\sqrt{2}|\beta|\cos(\Phi - \varphi)\right),$$

$$\nu = \frac{\Gamma \cdot \lambda\tau}{\sqrt{2}}(1+\mu), \qquad (12)$$

$$-1 \leq \mu \leq 1.$$

This approximation is essentially used in further calculations. Here $\mu$ is the fitting parameter, with which the average probability $\overline{P}_{10}$ varies in the interval

$$\frac{1}{2} \leq \overline{P}_{10} \leq P_{1,0}^{\beta}(t_1).$$

The graph 2, Fig.4, shows an example of such a line for $\mu = 0.76$. The approximation (12) greatly simplifies the analysis, as it allows applying the Bernoulli formula to calculation of integrated probability. We write this probability

$$P_{\beta}(n;m) = \sum_{j=1} P_{\beta}^{j}(n;m) \approx \frac{n!}{(n-m)!m!}\left(\overline{P}_{10}\right)^m \left(1-\overline{P}_{10}\right)^{n-m}.$$

Using local Moivre-Laplace theorem, we obtain the relation

$$P_{\beta}(n;m) \approx \frac{1}{\sqrt{2\pi n \overline{P}_{10}(1-\overline{P}_{10})}} \exp\left(-\frac{1}{2}\frac{(m-n\overline{P}_{10})^2}{n\overline{P}_{10}(1-\overline{P}_{10})}\right). \qquad (13)$$

In this formula, the random variable $m$ is changed in the interval $0 \leq m \leq n$. We define the distribution function $F_{\beta}(n;m)$ of a random variable $m$

$$F_{\beta}(n;x) = \sum_{m=0}^{x} P_{\beta}(n;m). \qquad (14)$$

Sample distribution function of a random value $m$ for the coherent state $|\beta\rangle$ is denoted $\mathrm{F}_{\beta}(n;x)$. To determine $\mathrm{F}_{\beta}(n;x)$, let us introduce the interval function with conditional operations

$$I(X,x) = \begin{cases} 1 & \text{if } X \le x \\ 0 & \text{if } X > x \end{cases}.$$

Then

$$F_\beta(n;x) = \frac{1}{N}\sum_{k=1}^{N} I(m_k,x), \qquad (15)$$

where $m_k$ $k=1...N$ is the sample value of a random variable $m$, $N$ is the sample size. Denote by $|\beta|_{max}$ - the maximum value of the amplitude of the coherent state, which defines the area of integration in equation (4). We choose the fitting parameter $\mu$ (12), graphically comparing the theoretical formula for $F_\beta(n;m)$ (14) with sample formula $F_\beta(n;x)$ (15). We will estimate quality of coincidence of functions of distribution by Kolmogorov's criterion [42] for confidence level $\alpha = 0.95$. According to this criterion, the confidence interval is given by

$$\max_{-\infty<x<\infty}|F_\beta(n;x) - F_\beta(n;x)| < \frac{\sqrt{\frac{1}{2}\ln\left(\frac{2}{1-0.95}\right)}}{\sqrt{N}}.$$

The simulation results of the random quantity $m$ for the sample size $N = 1000$ shown in Fig. 5. Choose from the following calculation parameters: $\lambda\tau = 0.04$, $n = 300$, $\mu = 0.76$, $\alpha = 0.95$, $|\beta|_{max} = 3$, $\varphi = -\frac{3\pi}{4}$, $\Phi = \frac{\pi}{4}$. Graph 1 shows a sample function $F_\beta(n;x)$, Graph 2 is Kolmogorov level, Graph 3 is norm of the difference of two functions. As can be seen from Fig. 5, Kolmogorov's criterion is reliably performed when selecting the fitting parameter $\mu = 0.76$. The simulation results show that the criterion is performed for all smaller values $|\beta| < |\beta|_{max}$.

## 4. Distribution function of the quadratures in the coherent state

The quadrature $\chi(\varphi)$ (1) is a random variable in a mode coherent state. We associate the random variable $\chi$ with a random value $m$ according to the formula

$$\chi = \frac{2m-n}{n\nu}. \qquad (16)$$

This is the main formula of our work. Let us rewrite formula (13) for the new random variable (16) and obtain

$$P_\beta(n;m) \to P_\beta(\chi(\varphi)) = \frac{1}{\sqrt{2\pi\sigma}}\exp\left(-\frac{1}{2}\frac{\left(\chi - \sqrt{2}|\beta|\cos(\Phi-\varphi)\right)^2}{\sigma}\right). \qquad (17)$$

Here we have used the approximation (11). Compare formulas (3) and (17). Equation (17) gives the normal distribution with the correct mean $\sqrt{2}|\beta|\cos(\Phi-\varphi)$ and variance

$$\sigma = \frac{1}{n\nu^2}, \qquad (18)$$

where the parameter $\nu$ defined by the formula (12). Fig. 5 is performed for the parameter values $\lambda\tau = 0.04$, $n = 300$, $\mu = 0.76$, $|\beta|_{max} = 3$, to which the variance (18) is equal to

$$\sigma = \frac{1}{nv^2} = 1.345.$$

The variance of the quadrature normal distribution in the formula (3) are equal $\frac{1}{2}$. The reason for this discrepancy is that the formula (17) describe the convolution of the true quadrature distribution (3) with the instrumental function $Spr(\chi)$ of our method

$$Spr(\chi) = \frac{1}{\sqrt{2\pi\sigma_S}} \exp\left(-\frac{1}{2}\frac{\chi^2}{\sigma_S}\right), \qquad (19)$$

which should be chosen in the form of normal distribution with a zero average and variance

$$\sigma_S = \sigma - \frac{1}{2}. \qquad (20)$$

We define the convolution of the theoretical quadrature distribution in the coherent state $P_\beta(\chi(\varphi))$ (3) and instrumental function (19)

$$Cnv_\beta(\chi) = \int_{-\infty}^{\infty} Spr(x) P_\beta(\chi - x) dx. \qquad (21)$$

We define a theoretical distribution function of coherent state quadrature

$$F_\beta(\chi) = \int_{-\infty}^{\chi} P_\beta(\chi) d\chi.$$

We define a theoretical distribution function of the convolution (21)

$$FC_\beta(\chi) = \int_{-\infty}^{\chi} Cnv_\beta(\chi) d\chi. \qquad (22)$$

Results of numerical simulation of a sample distribution function $FC_\beta(\chi)$ are presented in Fig.6. The variable $\chi$ associated with $m$ by the equation (16). The function $FC_\beta(\chi)$ is calculated by the formula

$$FC_\beta(x) = \frac{1}{N}\sum_{k=1}^{N} I(\chi_k, x),$$

$$\chi_k = \frac{2m_k - n}{nv}, \; k = 1...N.$$

Fig. 6 is calculated for fitting parameter $\mu = 0.76$, obtained for Fig. 5. As follows from Fig. 6, sample distribution function $FC_\beta(x)$ within the Kolmogorov confidence interval coincides with the theoretical distribution function $FC_\beta(\chi)$. Theoretical function $FC_\beta(\chi)$ obtained by convolution with the instrumental function (22).

## 5. The protocol detection of the quadrature distribution

The protocol to measure the quadrature distribution (1) in an arbitrary state $\rho_0$ of the microwave mode in the cavity consists of the following:

1. We estimate the range of variation $|\beta| \leq |\beta|_{max}$ for Glauber's distribution $P_0(\beta)$ in the studied state $\rho_0$;
2. We choose interaction time from a condition (11)
$$\beta_{max} \lambda \tau \ll 1; \qquad (23)$$
3. We select the number $n$ of atoms - probes in one experiment;
4. We simulate sample distribution function for the chosen parameters $|\beta| = |\beta|_{max}$, $n$, $\tau$ and select the fitting parameter $\mu$ for coincidence of sample $F_\beta(n;x)$ and theoretical $F_\beta(\chi)$ functions for Kolmogorov's criterion (Fig. 5);
5. We find the variance of the instrumental function $\sigma_S$ with formulas (20), (18), (12);
6. We carry out physical experiment of $N$ times on a setup of Fig. 1 with the found parameters $\tau, n, \mu$ for the studied state $\rho_0$. We define the sequence of sample values $m_k$, $k = 1....N$ of a random variable $m$. We calculate values of a random variable $\chi_k$, $k = 1....N$ on formulas (16), (12). We define sample distribution function of convolution of required quadrature distribution $P_{\rho_0}(\chi(\varphi))$ and the instrumental function $Spr(\chi)$;
7. We apply standard methods of deconvolution;

Let us verify the protocol on the example of the single-photon Fock state $|1\rangle$. Probability density of a quadrature for $|1\rangle$ is defined by a formula
$$P_{Fock}(\chi) = \frac{1}{2\sqrt{\pi}} \exp(-\chi^2) H_1(\chi)^2,$$
and doesn't depend on a phase $\varphi$. Here $H_n(\chi)$ is the Hermitic polynomial. Convolution of the density $P_{Fock}(\chi)$ and the instrumental function $Spr(\chi)$ (19) we define by the formula
$$Cnv_{Fock}(\chi) = \int_{-\infty}^{\infty} Spr(x) P_{Fock}(\chi - x) dx.$$

We introduce the notation of quadrature distribution function $F_{Fock}(\chi)$ in Fock state $|1\rangle$, the notation of distribution function $FC_{Fock}(\chi)$ of a convolution, the notation of sample distribution function $\text{FC}_{Fock}(\chi)$ of a convolution. As calculations show, two peaks on a function graph $P_{Fock}(\chi)$ don't differ at $\sigma_S > 0.8$. Rather narrow the instrumental function $Spr(x)$ is necessary for monitoring sharp dip between the peaks. Variance $\sigma_S = 0.845$ (20) was chosen at creation of Fig. 5 and Fig. 6. For receiving narrower instrumental function it is necessary to choose other set of parameters of experiment. Dependence of variance $\sigma_S$ of instrumental function on number $n$ at value $\lambda \tau = 0.04$ is shown in Fig. 7. Fig. 7 is constructed by analogy with Fig. 5 by selection of fitting parameter $\mu$ so that $\max_{-\infty < x < \infty} |F_\beta(n;x) - F_\beta(n;x)|$ got to the confidential interval determined by Kolmogorov's border. As appears from Fig. 7, at value of parameters $\lambda \tau = 0.04$, $|\beta|_{max} = 3$, $\varphi = -\frac{3\pi}{4}$, $\Phi = \frac{\pi}{4}$, $n = 1000$, $\mu = 0.36$ variance is equal

$$\sigma_s = 0.176.$$

We perform numerical simulations $\text{FC}_{Fock}(\chi)$ and present the results in Fig. 8. Parameters of simulation are: $\lambda\tau = 0.04$, $|\beta|_{max} = 3$, $\varphi = -\dfrac{3\pi}{4}$, $\Phi = \dfrac{\pi}{4}$, $n = 1000$, $\mu = 0.36$. As follows from Fig. 8, the above protocol reliably detects the convolution of the desired distribution function.

# 6. Discussion of results and conclusion

The protocol of measurement of quadrature distribution of microwave mode in CQED experiments was formulated in this work. According to idea of the protocol, the quantum tomogram of distribution $\chi(\varphi)$ for different values of a phase $\varphi$ can be received, measuring integrated probability $P_{\rho_0}(n;m)$ (8) of selective detectors counting on experimental setup Fig. 1. The integrated probability has to be measured at a right choice of parameters of experiment. The main parameter of the experiment is the interaction time $\tau$ of the atom - probe with a microwave mode. The range of variation of the argument $\beta$ of Glauber function $P_0(\beta)$ (4) is necessary to evaluate for a right choice of $\tau$. The parameter $\tau$ is calculated according to the relation (23), after evaluating $|\beta|_{max}$. Number $n$ of the atoms - probes passed via the cavity in one experiment is the next chosen parameter. The fitting parameter $\mu$ must be choosing after selecting the above parameters. The choice of the parameter $\mu$ is carried out in the numerical experiment, by analogy with the simulation of Fig. 5. We must calculate the variance $\sigma_S$ of the instrumental function of the method by the formulas (20), (18), (12) after a choice of parameters $\tau$, $n$, $\mu$. If instrumental function does not "obscure" the studied effect, then, according to the protocol, it is necessary to receive the sample distribution function of a quadrature. The above mentioned paragraphs of the Protocol are preliminary, needed for correct selection of setup parameters Fig.1. After the preparatory stage, we must carry out the physical experiments on the setup of Fig.1. Physical experiment is carried out $N$ times, as a result we receive sequence of a random variable of $m_k$, $k = 1....N$. Then it is necessary to calculate sample values of a quadrature on a formula (16) and to construct sample distribution function of convolution of the studied distribution and instrumental function $Spr(\chi)$ (19). For receiving required distribution it is necessary to perform a procedure of the deconvolution. A deconvolution operation is well-studied in the theory of digital signal processing. Features of a deconvolution protocol described for example in [43].

## Acknowledgements

This work was partially financially supported by the Government of the Russian Federation (grant 074-U01), by Ministry of Education and Science of the Russian Federation (GOSZADANIE 2014/190, Project14.Z50.31.0031 and ZADANIE No. 1.754.2014/K), by grant of Russian Foundation for Basic Researches and grant of the President of Russia (MK-2736.2015.2).

# Abstract

The protocol of measurement the probability density $P_\rho(\chi)$ of a rotated quadrature $\chi$ of microwave mode in an arbitrary quantum state is presented and numerically investigated. The protocol is adapted to the measuring procedure used in the CQED experiments. The method is based on measurement of integrated probability $P(m,n)$ of detection of $m$ atoms - probes in the top working state from $n$ atoms flying via the cavity. Connection of random variables $m = m(\varphi)$ and $\chi = \chi(\varphi)$ is found, when $\varphi$ is a qudrature phase. Connection of the tomogram $P(m,n)$ and the tomogram $P_\rho(\chi)$ is found. The protocol prescribes the rule of choice of problem parameters, such as interaction time, and the number of atoms in series $n$. The method contains a single fitting parameter that is selected in the process of numerical simulation of the quadrature distribution in a coherent state. Quality selection is determined by the Kolmogorov test. The result of the calculation protocol is a convolution $P_\rho(\chi)$ with the instrumental function of the method. The parameters of the instrumental function are shown. Numerical simulations performed for typical numerical values of parameters of the experimental setup, shows the effectiveness of the proposed method.

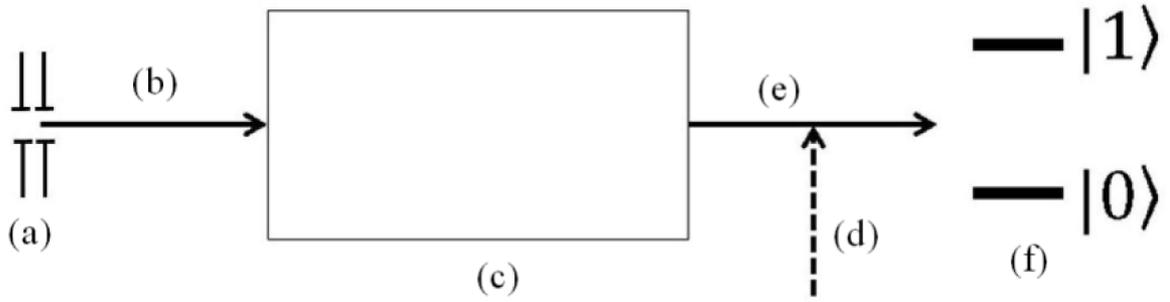

Fig.1. Diagram of the CQED setup for a tomography of quadratures distribution of microwave mode. a) The source of velocity selected rubidium $^{85}Rb$ Rydberg atoms in the main working state. b) The sequence of atoms-probes entering the cavity. c) High-Q microwave cavity containing photons of mode in studied state $\rho_0$. d) Microwave pulse, turning atomic basis and defining the phase $\varphi$ of the studying quadrature. e) The sequence of atoms-probes leaving the cavity. f) Selective detector of energy states of the atom-probe.

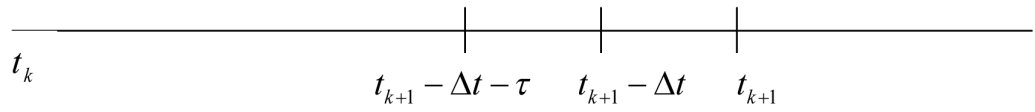

Fig. 2. The timing diagram of the protocol events. The time interval between the points $t_k$ and $t_{k+1}$ of measurements of the atom-probe states. $\Delta t$ is the pulse duration of the microwave field, turning an atomic basis at an angle $\dfrac{\pi}{4}$. $\tau$ is the duration of the interaction interval of the atom - probe with a microwave field.

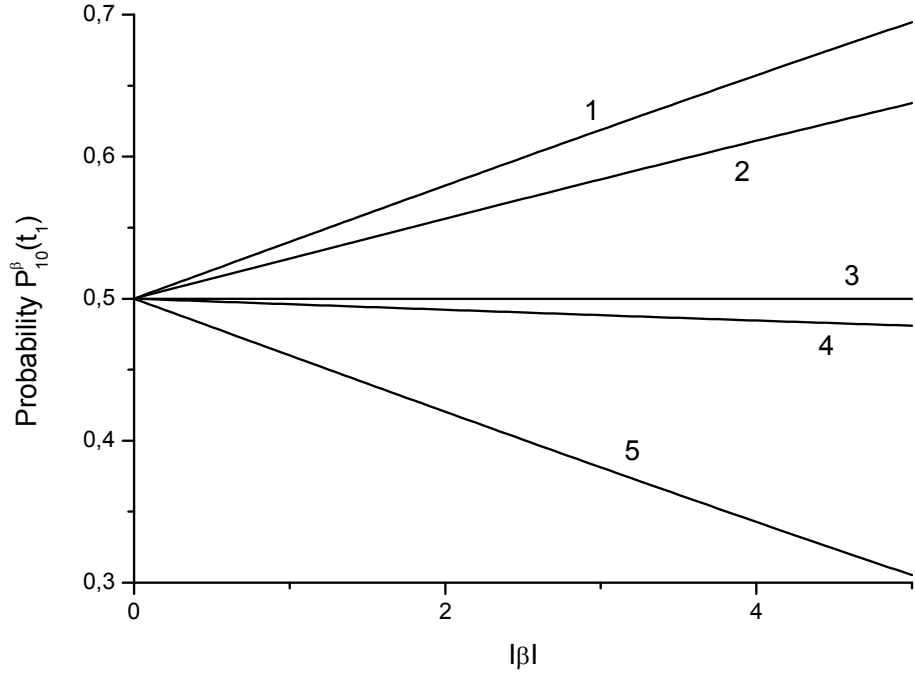

Fig.3. The a priori probability $P_{1,0}^{\beta}(t_1)$ of finding the atom - probe in the top state on the first measurement. The initial state of the atom is lower. Mode in the coherent state $|\beta\rangle\langle\beta|$. Amplitude $\beta = |\beta|\exp(i\Phi)$. Quadrature phase $\varphi$. Interaction parameter $\lambda\tau = 0.04$. Parameters of calculation: Graph 1 - $\Phi - \varphi = 0$, Graph 2 - $\Phi - \varphi = \dfrac{\pi}{4}$, Graph 3 - $\Phi - \varphi = \dfrac{\pi}{2}$, Graph 4 - $\Phi - \varphi = \dfrac{17}{32}\pi$, Graph 5 - $\Phi - \varphi = \pi$.

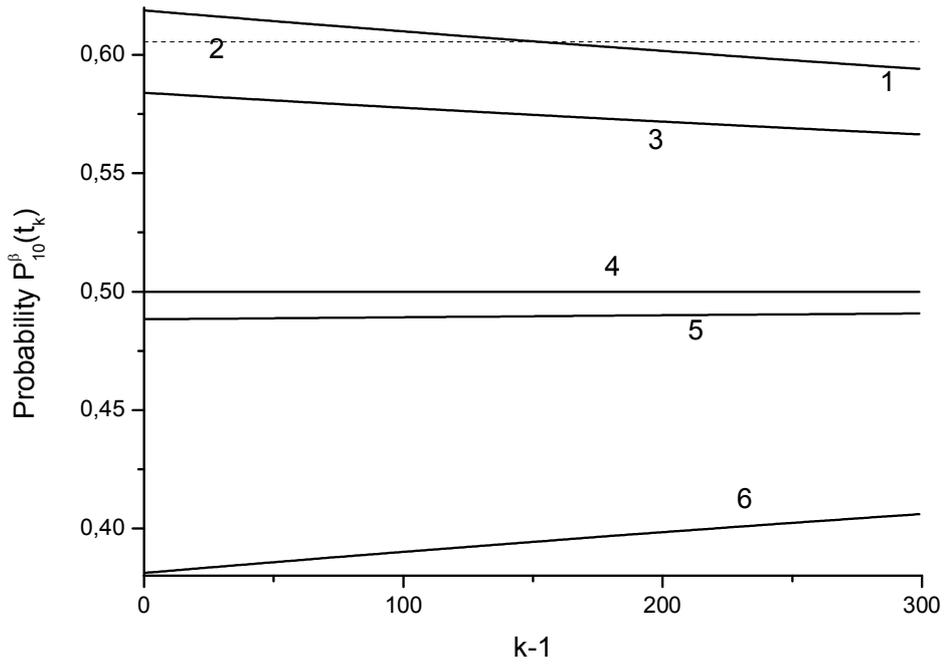

Fig. 4. The a priori probability of detecting $k$ - th atom-probe in an top state. Initial mode state - $|\beta\rangle\langle\beta|$. Graphs 1, 3-6 – probability calculation is made using the reduced density matrix obtained in each step in the sequence of = 300 flown atoms. Graph 2 – the average value $\overline{P}_{10}$, для $\mu = 0.76$ (12) . Parameters of calculation: $|\beta| = 3$, $\lambda\tau = 0.04$, Graph 1,2 - $\varphi + \Phi = 0$, Graph 3 - $\Phi - \varphi = \dfrac{\pi}{4}$, Graph 4 - $\Phi - \varphi = \dfrac{\pi}{2}$, Graph 5 - $\Phi - \varphi = \dfrac{17}{32}\pi$, Graph 6 - $\Phi - \varphi = \pi$

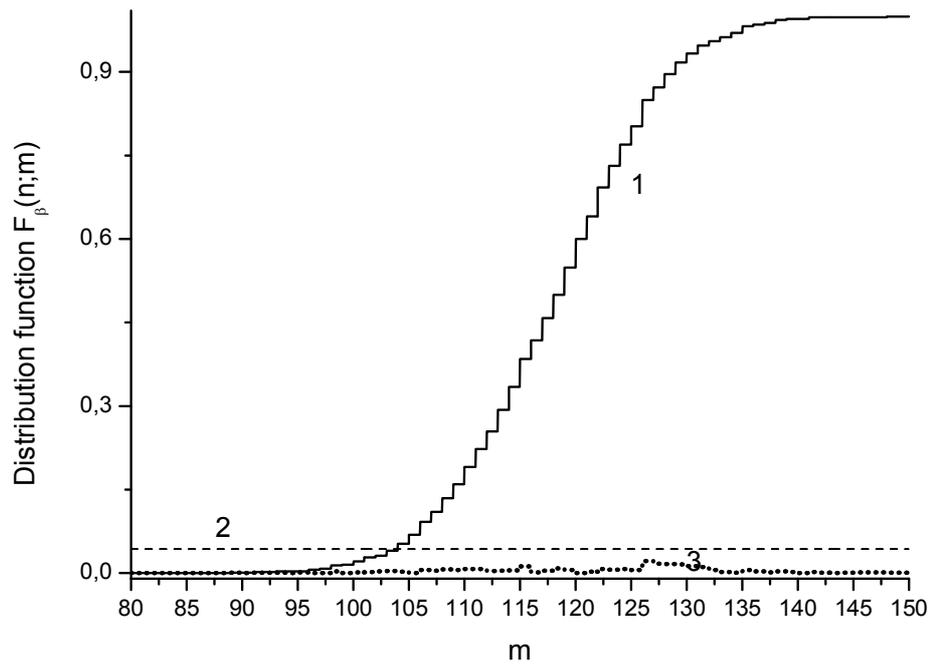

Fig. 5. Sample distribution function of a random variable $m$. Graph 1 – Sample distribution function $F_\beta(n;x)$, Graph 2 – Kolmogorov's level, Graph 3 – norm of a difference $\max\limits_{-\infty<x<\infty}|F_\beta(n;x)-F_\beta(n;x)|$. Parameters of calculation: $\lambda\tau=0.04$, $n=300$, $\mu=0.76$, $\alpha=0.95$, $|\beta|_{\max}=3$, $\varphi=-\dfrac{3\pi}{4}$, $\Phi=\dfrac{\pi}{4}$, $N=1000$.

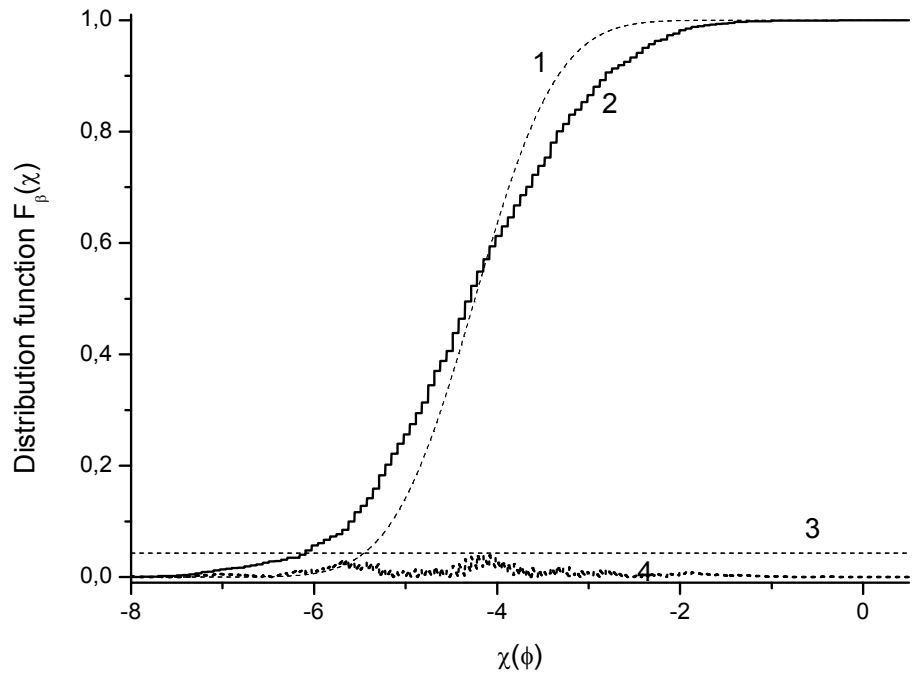

Fig. 6. Quadrature distribution function $\chi(\varphi)$ in a coherent state $|\beta\rangle$. Graph 1 – theoretical quadrature distribution function in a coherent state $F_\beta(\chi)$. Graph 2 – sample quadrature distribution function $\mathrm{FC}_\beta(x)$, Graph 3 – Kolmogorov's level of confidence. Graph 4 - norm of a difference $\max\limits_{-\infty<x<\infty}|\mathrm{FC}_\beta(x)-FC_\beta(x)|$. Parameters of calculation the same as for Fig.5.

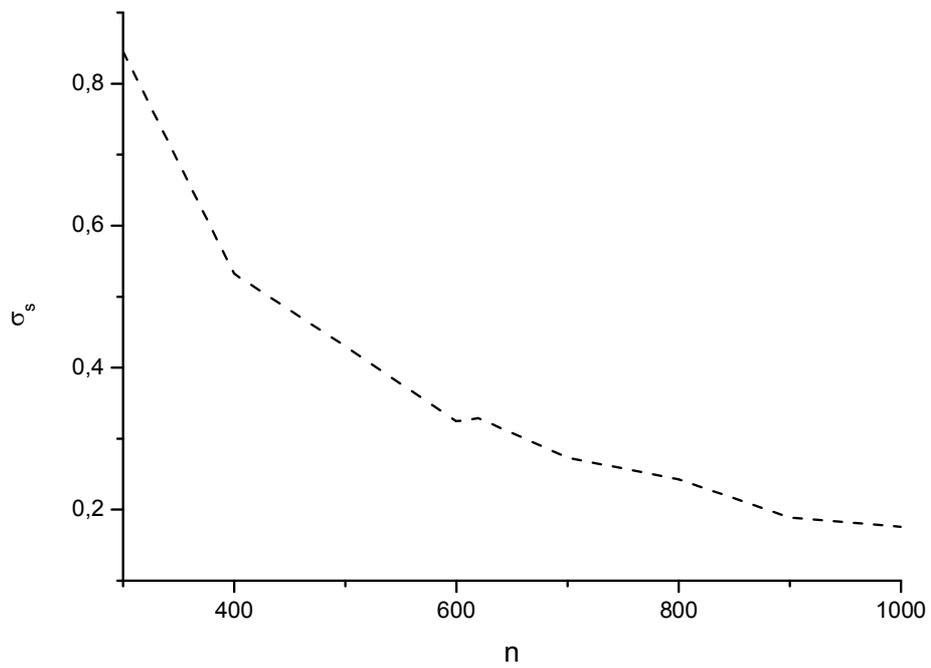

Fig. 7. The dependence of the variance of the instrumental function on the parameter $n$ at the parameters: $\lambda\tau = 0.04$, $\alpha = 0.95$, $|\beta|_{max} = 3$, $\varphi = -\dfrac{3\pi}{4}$, $\Phi = \dfrac{\pi}{4}$, $N = 1000$

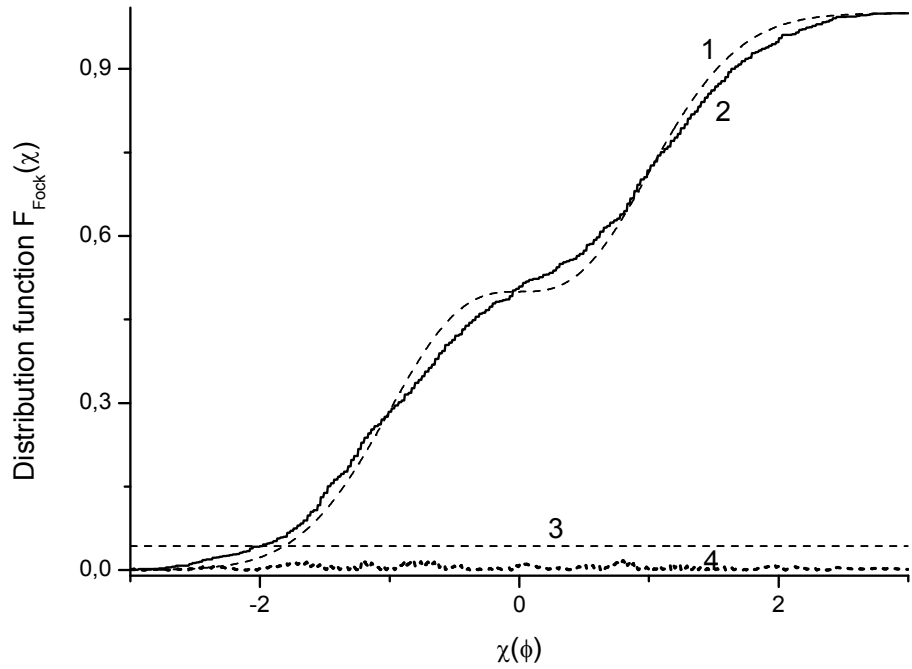

Fig. 8. Quadrature distribution function $\chi(\varphi)$ in one-photon state $|1\rangle$. Graph 1 – theoretical quadrature distribution function $F_{Fock}(\chi)$. Graph 2 – sample quadrature distribution function of convolution $FC_{Fock}(\chi)$, Graph 3 – Kolmogorov's level of confidence. Graph 4 - norm of a difference $\max_{-\infty<x<\infty}|FC_{Fock}(\chi)-FC_{Fock}(\chi)|$. Parameters of calculation: $\alpha=0.95$, $\mu=0.36$, $\lambda\tau=0.04$, $n=1000$, $N=1000$.

# Literature

[1] D'Ariano, G.M., Leonhardt, U., Paul, H.: Homodyne detection of the density matrix of the radiation field. Phys. Rev. A 52, R1801-R1804 (1995)

[2] Smithey, D.T., Beck, M., Raymer, M.G., Faridani, A.: Measurement of the Wigner Distribution and the Density Matrix of a Light Mode Using Optical Homodyne Tomography: Application to Squeezed States and the Vacuum. Phys. Rev. Lett. 70, 1244-1247 (1993)

[3] Smithey, D.T., Beck, M., Cooper, J., Raymer, M.G.: Measurement of number-phase uncertainty relations of optical fields. Phys. Rev. 48, 3159-3167 (1993)

[4] Raymer, M.G., Beck, M., McAlister, D.F.: Complex Wave-Field Reconstruction Using Phase-Space Tomography. Phys. Rev. Lett. 72, 1137-1140 (1994)

[5] Leonhardt, U., Paul, H.: Measuring the quantum state of light. Prog. Quant. Electr. 19, 89-130 (1995)

[6] Filippov, Sergey N., Man'ko, Vladimir I.: Measuring microwave quantum states: Tomogram and moments. Phys. Rev. A 84, 033827 (2011)

[7] Richter, Th.: Determination of field correlation functions from measured quadrature component distributions. Phys. Rev. A 53, 1197-1199 (1996)

[8] Wunsche, Alfred.: Tomographic reconstruction of the density operator from its normally ordered moments. Phys. Rev. A 54, 5291-5294 (1996)

[9] Dodonov, V. V.: Purity- and entropy-bounded uncertainty relations for mixed quantum states. J. Opt. B: Quantum Semiclass. Opt. 4, S98-S108 (2002)

[10] Man'ko, V.I., Marmo, G., Porzio, A., Solimeno, S., Ventriglia, F.: Homodyne estimation of quantum state purity by exploiting the covariant uncertainty relation. Phys. Scr. 83, 045001 (2011)

[11] Raimond, J. M., Brune, M., Haroche, S.: Colloquium: Manipulating quantum entanglement with atoms and photons in a cavity. Rev. Mod. Phys.73, 565-582 (2001)

[12] Blais, Alexandre, Ren-Shou Huang, Wallraff Andreas, Girvin, S. M., Schoelkopf, R.J.: Cavity quantum electrodynamics for superconducting electrical circuits: An architecture for quantum computation. Phys. Rev. A 69, 062320 (2004)

[13] Zurek, W.: Decoherence, einselection, and the quantum origins of the classical. Rev. Mod. Phys. 75, 715–775 (2003).

[14] Devoret, Michel H., Martinis, John M.: Implementing Qubits with Superconducting Integrated Circuits. Quantum Information Processing. 3, 163-203 (2004)

[15] Wallraff, A., Schuster, D.I., Blais, A., Frunzio, L., Huang, R.- S., Majer, J., Kumar, S., Girvin, S.M., Schoelkopf, R.J.: Strong coupling of a single photon to a superconducting qubit using circuit quantum electrodynamics. Nature. 431, 162-167 (2004)

[16] Hofheinz, Max, Wang, H., Ansmann, M., Bialczak, Radoslaw C., Lucero, Erik, Neeley, M., O'Connell, A.D., Sank, D., Wenner, J., Martinis, John M., Cleland, A. N.: Synthesizing arbitrary quantum states in a superconducting resonator. Nature. 459, 546-549(2009)

[17] Hofheinz, Max, Weig, E.M., Ansmann, M., Bialczak, Radoslaw C., Lucero, Erik, Neeley, M., O'Connell Wang, H., Martinis, John M., Cleland, A.N.: Generation of Fock states in a superconducting quantum circuit. Nature. 454, 310-314 (2008)

[18] Chen, Y.-F., Hover, D., Sendelbach, S., Maurer, L., Merkel, S.T., Pritchett, E.J., Wilhelm, F. K., McDermott, R.: Microwave Photon Counter Based on Josephson Junctions. Phys. Rev. Lett. 107, 217401 (2011)

[19] Koshino, Kazuki, Zhirong Lin, Kunihiro Inomata, Tsuyoshi Yamamoto, Yasunobu Nakamura.: Dressed-state engineering for continuous detection of itinerant microwave photons. arXiv:1509.05858 [quant-ph]

[20] Helmer, Ferdinand, Mariantoni, Matteo, Enrique Solano, Florian Marquardt.: Quantum nondemolition photon detection in circuit QED and the quantum Zeno effect. Phys. Rev. 79, 052115 (2009)

[21] Guillermo Romero, Juan José García-Ripoll, Enrique Solano.: Photodetection of propagating quantum microwaves in circuit QED. Phys. Scr. T137, 014004 (2009)
[22] Romero, G., Garcı́a-Ripoll, J.J., Solano, E.: Microwave Photon Detector in Circuit QED. Phys. Rev. Lett. 102, 173602 (2009)
[23] Mallet, F., Castellanos-Beltran, M.A., Ku, H.S., Glancy, S., Knill, E., Irwin, K.D., Hilton, G.C., Vale, L.R., Lehnert, K.W.: Quantum State Tomography of an Itinerant Squeezed Microwave Field. Phys. Rev. Lett. 106, 220502 (2011)
[24] Eichler, C., Bozyigit, D., Lang, C., Steffen, L., Fink, J., Wallraff, A.: Experimental State Tomography of Itinerant Single Microwave Photons. Phys. Rev. Lett. 106, 220503 (2011)
[25] Eichler, C., Bozyigit, D., Lang, C., Baur, M., Steffen, L., Fink, J.M., Filipp, S., Wallraff, A.: Observation of Two-Mode Squeezing in the Microwave Frequency Domain. Phys. Rev. Lett. 107, 113601 (2011)
[26] Bodendorf, C.T., Antesberger, G., Kim, M.S., Walther, H.: Quantum-state reconstruction in the one-atom maser. Phys. Rev. A. 57, 1371-1378 (1998)
[27] Bardroff, P.J., Mayr, E., Schleich, W.P., Domokos, P., Brune, M., Raimond, J.M., Haroche, S.: Simulation of quantum-state endoscopy. Phys. Rev. A. 53, 2736-2741 (1996)
[28] Martin Wilkens, Pierre Meystre.: Nonlinear atomic homodyne detection: A technique to detect macroscopic superpositions in a micromaser. Phys. Rev. A. 43, 3832-3835 (1991)
[29] Brune, M., Haroche, S., Raimond, J.M., Davidovich, L., Zagury, N.: Manipulation of photons in a cavity by dispersive atom-field coupling: Quantum-nondemolition measurements and generation of "Schrodinger cat" states. Phys. Rev. A. 45, 5193-5214 (1992)
[30] Lutterbach, L.G., Davidovich, L.: Method for Direct Measurement of the Wigner Function in Cavity QED and Ion Traps. Phys. Rev. Lett. 78, 2547-2550 (1997)
[31] Samuel Deleglise, Igor Dotsenko, Clement Sayrin, Julien Bernu, Michel Brune, Jean-Michel Raimond, Serge Haroche.: Reconstruction of non-classical cavity field states with snapshots of their decoherence. Nature. 455, 510-514 (2008)
[32] Clement Sayrin, Igor Dotsenko, Xingxing Zhou, Bruno Peaudecerf, Theo Rybarczyk, Sebastien Gleyzes, Pierre Rouchon, Mazyar Mirrahimi, Hadis Amini, Michel Brune, Jean-Michel Raimond, Serge Haroche.: Real-time quantum feedback prepares and stabilizes photon number states. Nature. 477, 73-77 (2011)
[33] Rybarczyk, T., Gerlich, S., Peaudecerf, B., Penasa, M., Julsgaard, B., Mølmer, K., Gleyzes, S., Brune, M., Raimond, J.M., Haroche, S., Dotsenko, I.: Past quantum state analysis of the photon number evolution in a cavity. arXiv:1409.0958v1 [quant-ph] 3 Sep 2014
[34] Mancini, S., Tombesi, P., Manko, V.I.: Density matrix from photon number tomography. Europhys. Lett. 37 (2), 79-83 (1997)
[35] Olga Man'ko, Man'ko, V.I.: Photon-Number tomography of multimode states and positivity of the density matrix. Journal of Russian Laser Research. 24, 5-12 (2003)
[36] Konrad Banaszek, Kryzysztof Wódkiewicz.: Direct Probing of Quantum Phase Space by Photon Counting. Phys. Rev. Lett. 76, 4344-4347 (1996)
[37] Roy Glauber.: Coherent and Incoherent States of the Radiation Field. Phys. Rev. 131, 2766-2788 (1963)
[38] Cahill, K.E., Glauber, R.J.: Density Operators and Quasiprobability Distributions. Phys. Rev. 177, 1882-1902 (1969)
[39] Miroshnichenko, G. P.: Discrete Photodetection of Quantum Jumps on the V Configuration of Atomic Levels. Journal of Experimental and Theoretical Physics. 109, 193–206 (2009)
[40] Miroshnichenko, G.P.: Phase Probability Density in Subensembles of a Quantum Mode of a One-Atom Maser. Optics and Spectroscopy. 97, 403–410 (2004)
[41] Miroshnichenko, G.P.: Statistics of Discrete Photodetection of Resonance Fluorescence in the Mollow Sidebands. Journal of Experimental and Theoretical Physics. 107, 952–959 (2008)
[42] Borovkov, A.A.: Mathematical statistics. Gordon and Breach Science Publishers (1998).
[43] Emmanuel Ifeachor, Barrie Jervis.: Digital Signal Processing: A Practical Approach. Addison-Wesley-Publishing Company (1993)